\documentclass[showpacs,preprintnumbers,amsmath,amssymb,onecolumn,12pt]{revtex4-1}
\usepackage{graphicx}
\usepackage{dcolumn}
\usepackage{bm}
\usepackage{subfigure}
\usepackage{float}
\usepackage{multirow}
\usepackage{booktabs}
\usepackage{amsmath}
\usepackage{cases}
\usepackage{latexsym,amssymb}
\usepackage[mathscr]{eucal}
\usepackage{lineno}
\usepackage{physics}
\usepackage[bookmarks=true,
   colorlinks=true,
   linkcolor=blue,
   urlcolor=blue,
   citecolor=blue,
   bookmarks=true,
   hyperindex=true
]{hyperref}
\providecommand{\U}[1]{\protect\rule{.1in}{.1in}}

\newcommand{\f}{\begin{equation}}
\newcommand{\ff}{\end{equation}}
\newcommand{\fa}{\begin{eqnarray}}
\newcommand{\ffa}{\end{eqnarray}}

\begin{document}
\allowdisplaybreaks[4]
\title{Spherical photon orbits around Kerr-MOG black hole}
\author{Shi-Yu Li $^{1}$}
\author{Song-Shan Luo$^{1}$}
\author{Zhong-Wen Feng$^{1}$}
\email{zwfengphy@163.com}
\affiliation{$^1$ School of Physics and Astronomy, China West Normal University, Nanchong 637009, China}

\begin{abstract}
This study investigates photon orbits around Kerr-MOG black holes. The equation of photon of motion around the Kerr-MOG black hole is derived by solving the Hamilton-Jacobi equation, expressed as a sixth-order polynomial involving the inclination angle $v$, the rotation parameter $u$, and the deformation parameter $\alpha$ that characterizes modified gravity. We find that $\alpha$ constrains the rotation of the black hole, modifying its gravitational field and leading to distinct photon orbital characteristics. Numerical analysis reveals that the polar plane ($v=1$) has two effective orbits: one outside and one inside the event horizon, while the equatorial plane ($v=0$) has four effective orbits: two outside and two inside the event horizon. Moreover, we derive the exact formula for general photon orbits between the polar and equatorial planes ($0<v<1$). In the extremal case, the rotation speed significantly impacts general photon orbits. A slowly rotating extremal black hole has two general photon orbits outside the event horizon, whereas a rapidly rotating extremal black hole has only one such orbit. In the non-extremal case, a critical inclination angle $v_{cr}$ exists in the parameter space $\left(v, u, \alpha \right)$. Below $v_{cr}$, there are four general photon orbits, while above $v_{cr}$, there are two orbits. At the critical inclination angle, three solutions are found: two photon orbits outside and one inside the event horizon. Additionally, the results indicate that all orbits are radially unstable. Furthermore, by analyzing photon impact parameter, we argue that $\alpha$ influences observational properties of the black hole.
\end{abstract}
\maketitle

\section{Introduction}
\label{intro}
Recent advancements in astronomical observation technologies have significantly enhanced our ability to study black holes in detail. FFor instance, the Event Horizon Telescope (EHT) captured the first image of a black hole in 2019, offering direct visual evidence of the supermassive black hole's shadow in M87 \cite{ref1,ref2,Akiyama2019c}. Additionally, gravitational wave observations by LIGO have opened new avenues for black hole research, particularly by detecting mergers of binary black hole systems. These advancements have revolutionized black hole research, allowing unprecedented observation and analysis of these enigmatic objects \cite{LIGOScientific:2016lio,KAGRA:2023pio}. Among the various aspects of black hole research, the study of photon orbits has emerged as a focus. The intense gravitational fields near black holes significantly affect photon paths, making their study essential for understanding phenomena such as gravitational lensing and black hole shadows. By investigating the behavior of photons near black holes, researchers can obtain the fundamental properties of these objects, including their mass, spacetime geometry, and gravitational effects. Such studies not only refine theoretical models but also provide essential tools for interpreting astronomical observations.

To the best of our knowledge, extensive studies on photon orbits around static black holes have provided significant insights. For example, in the Schwarzschild black hole, the photon sphere is a defining feature~\cite{Bardeen1973a,Chandrasekhar1985}. This region marks the smallest stable circular orbit for photons and plays a vital role in gravitational lensing and black hole shadow formation. Additionally, the presence of other substances or fields in spacetime imparts unique characteristics to photon spheres. These findings have established a solid foundation for understanding photon trajectories in static spacetimes. Despite significant progress in studying static black holes, it is widely acknowledged that most astrophysical black holes are rotating \cite{Lu:2023bbn,Cui:2023uyb}. Unlike the static black holes, rotating black holes like Kerr black holes display far more complex and diverse photon orbits due to their spin \cite{Carter1968,Hod2013}. Rotation induces frame-dragging effects, causing asymmetry in photon orbits and differentiating prograde from retrograde paths.	These features make the study of rotating black holes especially compelling and scientifically significant.

Current research on analytic formulas for photon orbits around rotating black holes has primarily focused on the equatorial and polar planes~\cite{Bardeen1972,bardeen1975lense,Teo2003}. This focus stems from the mathematical simplicity afforded by symmetry and its relevance to astronomical observations, as many astrophysical systems, such as accretion disks, align with the black hole’s equatorial region~\cite{Nealon:2015jya,Kumar:2020ltt,Islam:2021ful,Galison:2024bop}. While these studies have provided significant insights, real-world observations often involve detectors at various angles relative to the black hole~\cite{Chatterjee:2020eqc}, necessitating investigations beyond the equatorial and polar planes~\cite{Yang:2012he,Hod2013}. Recently, Tavlayan and Tekin~\cite{Tavlayan2020} provided valuable insights into the exact formulas for photon orbits between the equatorial and polar planes of Kerr black hole. Their work reveals a critical inclination angle, below which two distinct photon orbits exist outside the event horizon.	At the critical inclination angle, the radius of the photon orbit around a black hole can be determined analytically. Building on this foundation, subsequent studies have explored photon orbits in rotating black holes with charge, perfect dark fluid matter, quintessence and electromagnetic field, revealing the additional influence of electric charge, dark matter, electromagnetic field, or high-dimensional spacetime on these orbits \cite{Cao:2022bvu,Fathi:2022ntj,Tavlayan:2022hzl,Chen2023,Alam:2024mmw}. Collectively, these findings enhance our understanding of photon behavior in rotating spacetimes, providing crucial insights for interpreting their observable properties.	

On the other hand, general relativity (GR), the cornerstone of modern black hole research, faces challenges in explaining certain astrophysical phenomena, such as galaxy rotation curves \cite{Rodrigues:2009vf,Ciotti:2024uvo}, galaxy cluster properties \cite{Geller:2012ds,Hogan:2016pmv}, and the stability of galactic spiral arms \cite{tenjes2017spiral}. However, despite extensive efforts, direct detection of dark matter remains inconclusive. This challenge has motivated researchers to explore alternative gravitational theories, among which scalar-tensor-vector gravity (STVG), also known as modified gravity (MOG), stands out \cite{Moffat2005,Brownstein:2007sr,Moffat:2016gkd}. By introducing additional scalar, tensor, and vector fields, MOG modifies gravitational interactions to address these anomalies without invoking dark matter. The MOG framework aligns with observational data in various astrophysical contexts, providing a effective alternative for interpreting phenomena that challenge GR. Building on this framework, the Kerr-MOG black hole solution has emerged as an extension of the Kerr solution within the MOG theory \cite{Moffat2015a,Moffat2015b,Lee2017,Moffat2021,Moffat2021a,Liu2023}, introducing modifications to account for deviations from GR. Nowadays, Kerr-MOG black hole has become an essential tool for studying the impact of modified gravitational theory on phenomena like black hole thermodynamics, shadows, and gravitational lensing~\cite{Moffat2009,Moffat2015c,Mureika2016,Manfredi:2017xcv,Wei2018,Guo2018,Sheoran2018,Ovgun2019,Wang2019,Rahvar2019,Izmailov2019,Wang2019,Qiao2020,Liu:2024lbi}.

It should be noted that, despite the increasing interest in Kerr-MOG black holes, a comprehensive understanding of their photon orbits remains incomplete \cite{Lee2017,Cunha2017}. To address this issue, the present study analyzes photon orbits around Kerr-MOG black holes, encompassing not only polar and equatorial configurations but also intermediate regions. This research aims to characterize these photon orbits in detail, with particular emphasis on the influence of the deviation parameter $\alpha$ and its distinctions from conventional Kerr black holes. The structure of this paper is organized as follows. The section~\ref{sec2} provides a brief review of the Kerr-MOG black hole. In section~\ref{sec3}, starting from the Hamilton-Jacobi equation, we derive a sixth-order polynomial equation that relates the photon orbit radius around a Kerr-MOG black hole to the rotation parameter $u$, the deviation parameter $\alpha$, and the effective inclination angle $v$. The section~\ref{sec4} discusses photon orbits on the polar and equatorial planes, respectively. In section~\ref{sec5}, we extend the analysis to general photon orbits between the polar and equatorial planes. Finally, the conclusion and discussion are presented in Section~\ref{sec6}.

\section{Kerr-MOG black hole}
\label{sec2}
In this section, we briefly review the Kerr-MOG black hole. In Ref.~\cite{Moffat2005}, the action corresponding to STVG theory is given by
\begin{equation}
S = S_{\mathrm{GR}} + S_{\phi} + S_{s} + S_{\mathrm{M}},
\label{eq1}
\end{equation}
  with
\begin{equation}
S_{\mathrm{GR}} = \frac{1}{16\pi} \int \mathrm{d}^{4} x \sqrt{-g} \frac{R}{G},
\label{eq2}
\end{equation}

\begin{equation}
S_{\phi} = \int \mathrm{d}^{4} x \sqrt{-g} \left( -\frac{1}{4} B^{\mu\nu} B_{\mu\nu} + \frac{1}{2} \mu^{2} \phi^{\mu} \phi_{\mu} \right),
\label{eq3}
\end{equation}

\begin{equation}
S_{s} = \int \mathrm{d}^{4} x \sqrt{-g} \left[ \frac{1}{G^{3}} \left( \frac{1}{2} g^{\mu\nu} \nabla_{\mu} G \nabla_{\nu} G - V(G) \right) + \frac{1}{\mu^{2} G} \left( \frac{1}{2} g^{\mu\nu} \nabla_{\mu} \mu \nabla_{\nu} \mu - V(\mu) \right) \right],
\label{eq4}
\end{equation}
where $R$ is the Ricci scalar, $S_{\mathrm{GR}}$ and $ S_{\mathrm{M}}$ represent the action of Einstein's gravity and the matter, respectively. $ S_{s} $ is the action of a scalar field that is associated with potentials $V(G)$ and $V(\mu)$. Besides, the action $S_{\phi}$ describes a Proca field with mass $\mu$, and the tensor field $B_{\mu \nu}$ is constructed with the Proca-type massive vector field as
$B_{\mu \nu} = \partial_{\mu} \phi_{\nu} - \partial_{\nu} \phi_{\mu}$, which satisfies the following equations
\begin{equation}
{\nabla _\nu }{B^{\mu \nu }} = \frac{1}{{\sqrt { - g} }}{\partial _\nu }\left( {\sqrt { - g} {B^{\mu \nu }}} \right) = 0,
\label{eq5}
\end{equation}
\begin{equation}
\nabla_{\sigma}\mathrm{B}_{\mu\nu} + \nabla_{\mu}\mathrm{B}_{\nu\sigma} + \nabla_{\nu}\mathrm{B}_{\sigma\mu} = 0.
\label{eq6}
\end{equation}
Since the vector field mass $\mu$ has a negligible effect on the black hole solution and $G$ can be regarded as a constant independent of spacetime coordinates, the vacuum solution of this action simplifies to	
\begin{equation}
S=\int \mathrm{d}^4x\sqrt{-g}\left(\frac{R}{16\pi G}-\frac14B^{\mu\nu}B_{\mu\nu}\right).
\label{eq7}
\end{equation}
The field equation corresponding to action~(\ref{eq7}) is given by
\begin{equation}
G_{\mu\nu}=-8\pi GT_{\phi\mu\nu},
\label{eq8}
\end{equation}
where ${G_{\mu \nu }} = {R_{\mu \nu }} - {{R{g_{\mu \nu }}} \mathord{\left/ {\vphantom {{R{g_{\mu \nu }}} 2}} \right.  \kern-\nulldelimiterspace} 2}$ is Einstein tensor, and the energy momentum tensor of a vector field is ${T_{\phi \mu \nu }} =  - {{\left( {B_\mu ^\sigma {B_{\nu \sigma }} - {{{g_{\mu \nu }}{B^{\sigma \beta }}{B_{\sigma \beta }}} \mathord{\left/ {\vphantom {{{g_{\mu \nu }}{B^{\sigma \beta }}{B_{\sigma \beta }}} 4}} \right. \kern-\nulldelimiterspace} 4}} \right)} \mathord{\left/  {\vphantom {{\left( {B_\mu ^\sigma {B_{\nu \sigma }} - {{{g_{\mu \nu }}{B^{\sigma \beta }}{B_{\sigma \beta }}} \mathord{\left/ {\vphantom {{{g_{\mu \nu }}{B^{\sigma \beta }}{B_{\sigma \beta }}} 4}} \right.  \kern-\nulldelimiterspace} 4}} \right)} {4\pi }}} \right.  \kern-\nulldelimiterspace} {4\pi }}$. To facilitate the analysis, a dimensionless deformation parameter $\alpha$ ($\alpha > 0$) is introduced to quantify the deviation between MOG and GR. The enhanced gravitational constant $G$ is related to the Newtonian gravitational constant
$G_{\text{N}}$ by the relation $G_{\text{N}}$ can be expressed as $G = G_{\text{N}}\left(1 + \alpha \right)$. By solving Eq.~(\ref{eq8}) in the Boyer-Lindquist coordinates, the Kerr-MOG black hole metric is given by~\cite{Moffat2015a,Moffat2015b}
\begin{align}
{\text{d}}s^{2} = & -\frac{\Delta - a^2 \sin^2 \theta}{\rho^2} {\text{d}}t^2 + \sin^2 \theta \left[ \frac{(r^2 + a^2)^2 - \Delta a^2 \sin^2 \theta}{\rho^2} \right] {\text{d}}\phi^2
\nonumber \\
& - 2a \sin^2 \theta \left( \frac{r^2 + a^2 - \Delta}{\rho^2} \right) {\text{d}}t {\text{d}}\phi + \frac{\rho^2}{\Delta} dr^2 + \rho^2 {\text{d}}\theta^2,
\label{eq10}
\end{align}
with
\begin{equation}
\Delta  = {r^2} - 2G_{\text{N}} M r + {a^2} + \frac{\alpha }{{1 + \alpha }}G_{\text{N}}^2{M ^2}, \quad \rho^2 = r^2 + \alpha^2 \cos^2 \theta,
\label{eq11}
\end{equation}
where $M$ represents the Arnowitt-Deser-Misner (ADM) mass, related to the Newtonian mass $m$ as $M = m(1 + \alpha)$ \cite{Sheoran2018}, and $a$ denotes the angular momentum of black hole. It is straightforward to see that the horizons of a Kerr-MOG black hole are determined by the condition $\Delta = 0$, which gives
\begin{equation}
{r_ \pm } = M \pm \sqrt {\frac{{{M^2}}}{{1 + \alpha }} - {a^2}},
\label{eq13}
\end{equation}
where $r_{+}$ is the event horizon and $r_{-}$ is the Cauchy horizon, respectively. However, for $\alpha = 0$, the horizons reduce to those of the standard Kerr case. For simplicity, we set $G_{\text{N}}=1$ in subsequent calculations.

\section{Equations of motion in Kerr-MOG spacetime}
\label{sec3}
To study the photon orbits around the Kerr-MOG black hole, it is crucial to analyze the photon geodesics. These geodesics, which describe the motion of photons in the spacetime, are derived from the following Hamilton-Jacobi equation
\begin{equation}
\frac{\partial S}{\partial\lambda}=-\frac12g^{\mu\nu}\frac{\partial S}{\partial x^\mu}\frac{\partial S}{\partial x^\nu},
\label{eq14}
\end{equation}
where $\lambda$ is the affine parameter along the null geodesics, and $S$ is the Jacobi action. Since the black hole spacetime is axisymmetric, it possesses two Killing vector fields, $ \xi^t$ and $\xi^\phi$, associated with the photon energy $E$ and the $z$-component of the photon angular momentum $L_z$, respectively. Utilizing these Killing vector fields, the action $S$ can be expressed as~\cite{Moffat2006}
\begin{equation}
S =  - Et + {L_z}\phi  + {S_r}\left( r \right) + {S_\theta }\left( \theta  \right).
\label{eq15}
\end{equation}
By combining Eq.~(\ref{eq14}) with Eq.~(\ref{eq15}), the equations of motion of photons are given by \cite{Hod2013, Teo2003, Bardeen1972, Wilkins1972}
\begin{subequations}
\label{eq16}
\begin{align}
{\rho ^2}\frac{{\rm{d}}t}{{\rm{d}\lambda }}& = a\left( {{L_z} - aE{{\sin }^2}\vartheta } \right) + \frac{{{r^2} + {a^2}}}{\Delta }\left[ {\left( {{r^2} + {a^2}} \right)E - a{L_z}} \right],
\\
\rho^{2}\frac{\rm{d}\phi}{\rm{d}\lambda} &= \frac{{{L_z}}}{{{{\sin }^2}\vartheta }} - aE + \frac{a}{\Delta }\left[ {\left( {{r^2} + {a^2}} \right)E - a{L_z}} \right],
\\
{\rho ^2}\frac{{{\text{d}}r}}{{{\text{d}}\lambda }} &= \pm \sqrt {R\left( r \right)},
\\
\rho^{2}\frac{\rm{d}\theta}{\rm{d}\lambda} &= \pm\sqrt{\Theta \left(\vartheta \right)},
\end{align}
\end{subequations}
with explicit forms of the radial and angular equations
\begin{subequations}
\label{eq20}
\begin{align}
R\left( r \right) &= {\left[ {\left( {{r^2} + {a^2}} \right)E - a{L_z}} \right]^2} - \Delta \left[ {\mathcal{K} + {{\left( {{L_z} - aE} \right)}^2}} \right],
\\
\Theta \left(\vartheta \right) & = \mathcal{K} + {\cos ^2}\vartheta \left( {{a^2}{E^2} - \frac{{L_z^2}}{{{{\sin }^2}\vartheta }}} \right),
\end{align}
\end{subequations}
with the Carter constant $\mathcal{K} $, which is the third motion constant of the photon in the rotating black hole spacetime, affecting the motion of the photon in the latitude direction, and this conserved quantity is not related to any spacetime symmetry \cite{Carter1968}. To investigate spherical photon orbits, one must analyze the radial equation $R\left(r\right)$, which expands to yield the following expression:
\begin{align}
R \left(r\right) =  - \Delta \left[ {\mathcal{K} + {{\left( {{L_z} - aE} \right)}^2}} \right] + {\left( {{r^2} + {a^2}} \right)^2}{E^2} + {a^2}L_z^2- 2a\left( {{r^2} + {a^2}} \right)E{L_z}.
\label{eq22}
\end{align}
Obviously, when the deviation parameter $ \alpha = 0 $, the result returns to the Kerr case. In the radial direction, the photon orbits with constant radius $r$ are determined by two conditions $R\left( r \right) = 0$ and ${{{\text{d}}R\left( r \right)} \mathord{\left/
 {\vphantom {{{\text{d}}R\left( r \right)} {{\text{d}}r}}} \right.
 \kern-\nulldelimiterspace} {{\text{d}}r}} = 0$. Using the two conditions together with the Eq.~(\ref{eq22}), one yields two parameters as follows \cite{Teo2003, Bardeen1972}
\begin{subequations}
\label{eq23}
\begin{align}
\xi \left( r \right) & = \frac{L_z}{E}=\frac{a^2 M (1 + \alpha) + r \left[ M^2 \alpha - M r (1 + \alpha) + (1 + \alpha) \Delta \right]}{a (M - r) (1 + \alpha)}
,
\\
\eta \left( r \right) &= \frac{K}{{{E^2}}} = \frac{{{r^2}\left\{ {4Mr\left[ {{M^2}\alpha  + {a^2}\left( {1 + \alpha } \right)} \right] - 4\Delta {M^2}\alpha  - {r^2}{{\left( {r - 3M} \right)}^2}\left( {1 + \alpha } \right)} \right\}}}{{{a^2}{{\left( {M - r} \right)}^2}\left( {1 + \alpha } \right)}}.
\end{align}
\end{subequations}

To quantify the deviation between the photon orbit and the black hole's equatorial plane, an effective inclination angle is defined as follows~\cite{Hod2013}
\begin{equation}
\cos{i}=\frac{{L}_{z}}{\sqrt{{L}_{z}^2+\mathcal{K}}},
\label{eq25}
\end{equation}
which is a constant of motion.When the photon’s orbital plane coincides with the black hole’s equatorial plane, the effective inclination angle is $\pm 1$, and the Carter constant is $\mathcal{K} = 0$. In the polar plane of the black hole, the value of the effective inclination is $0$, in which case the photon angular momentum is $ L_z = 0$. For the sake of convenience, one can define the dimensionless variables as follows
\begin{equation}
x=\frac r{M},
\quad u=\frac{a^2}{M^2},
\quad v=\sin^2 i.
\label{eq26}
\end{equation}
In the following, we refer to $ u $ as the rotation parameter. The ranges satisfied by the above variables are:
\begin{equation}
x\geq 0,\quad 0 \leq u \leq 1/\left( 1+ \alpha \right),\quad 0\leq v\leq1.
\label{eq27}
\end{equation}
To avoid the black hole singularity, we derive the upper limit of $u$ from Eq.~(\ref{eq13}). When $\alpha = 0$, the spacetime reduces to the Kerr black hole, and the range of $u$ is $\left(0, 1\right)$. However, when considering the MOG effect, the upper limit of the rotation parameter is constrained by $\alpha$, meaning $u$ must be less than $1$.

According to the equivalence principle, photons with different energies can travel at the same radius, and their trajectories are determined solely by the gravitational field, independent of the photon energy. Therefore, combining Eq.(\ref{eq20}) with Eq.~(\ref{eq22}) results in a sextic polynomial~\cite{Bardeen1972, Wilkins1972, Hod2013}
\begin{align}
\label{eq28}
f(x) &= {\left( {1 + \alpha } \right)^2}{x^6} - 6{\left( {1 + \alpha } \right)^2}{x^5} + \left( {1 + \alpha } \right)\left( {2u v a + 13\alpha  + 9} \right){x^4}  \nonumber\\
&- 4\left( {1 + \alpha } \right)\left( {u\alpha  + 3\alpha  + u} \right){x^3} + \left[ {2u\left( {2\alpha  - 3v} \right)\left( {1 + \alpha } \right) + {u^2}v\left( {1 + {\alpha ^2}} \right)} \right.
\nonumber\\
&\left. { + 2\alpha uv\left( {u - 6} \right) + 4{\alpha ^2}} \right]{x^2} + 2uv\left( {1 + \alpha } \right)\left( {u\alpha  + 2\alpha  + u} \right)x + {u^2}v{\left( {1 + \alpha} \right)^2}.
\end{align}
Clearly, the roots of the polynomial represent the photon orbits around the black hole. However, solving the polynomial directly is challenging. Hence, we will analyze these orbits in the following sections based on three distinct cases.

\section{The photon orbits of polar and equatorial planes}
\label{sec4}
\subsection{The photon orbits of polar plane}
\label{sec4-1}
For polar orbits, one has $ L_z = 0 $ (i.e. $i =  \pm {\pi  \mathord{\left/ {\vphantom {\pi  2}} \right. \kern-\nulldelimiterspace} 2}$ or $ \nu = 1$). In this case, Eq.~(\ref{eq28}) reduces to a cubic polynomial as follows
\begin{equation}
f\left(x\right)=\left(1+\alpha\right)u+\left[2\alpha+\left(1+\alpha\right)u\right]x-3\left(1+\alpha\right)x^2+\left(1+\alpha \right)x^3,
\label{eq29}
\end{equation}
which give three real roots $x_1$, $x_2$, and $x_3$. Furthermore, the event horizon of the Kerr-MOG black hole can be expressed as follows:
\begin{equation}
x_h=1+\sqrt{1-u-\frac\alpha{1+\alpha}}.
\label{eq30}
\end{equation}

\vspace{-1.3\baselineskip}
\begin{figure}[htbp]
\centering
\includegraphics[width=0.65\textwidth]{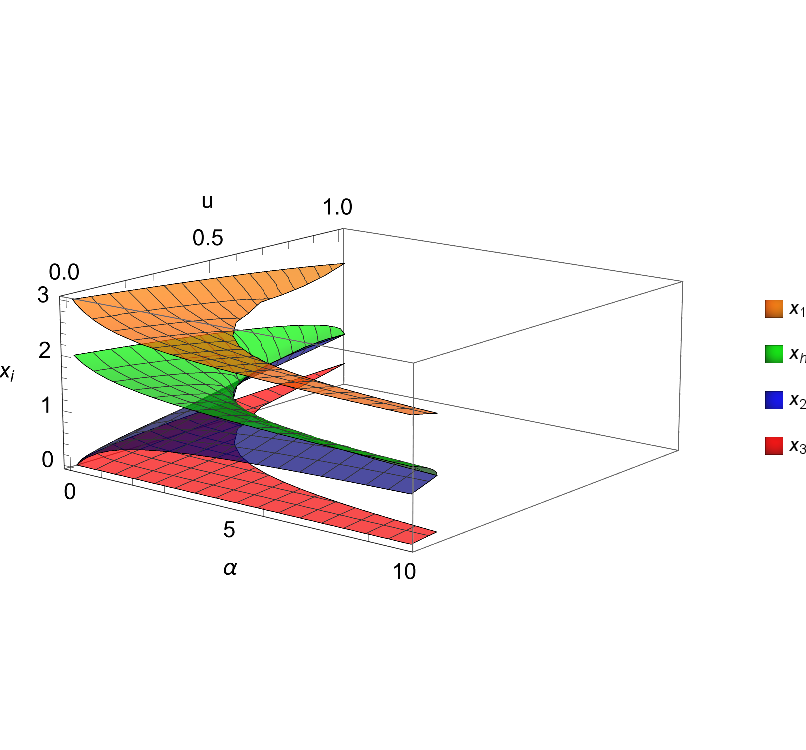}
\vspace{-2.5\baselineskip}
\caption{The polar photon orbits $x_i$ and event horizon $x_h$ of Kerr-MOG black hole as the functions of the rotation parameter $u$ and deviation parameter $\alpha$.}
\label{fig1}
\end{figure}
\begin{figure}[htbp]
\centering
\includegraphics[width=0.7\textwidth]{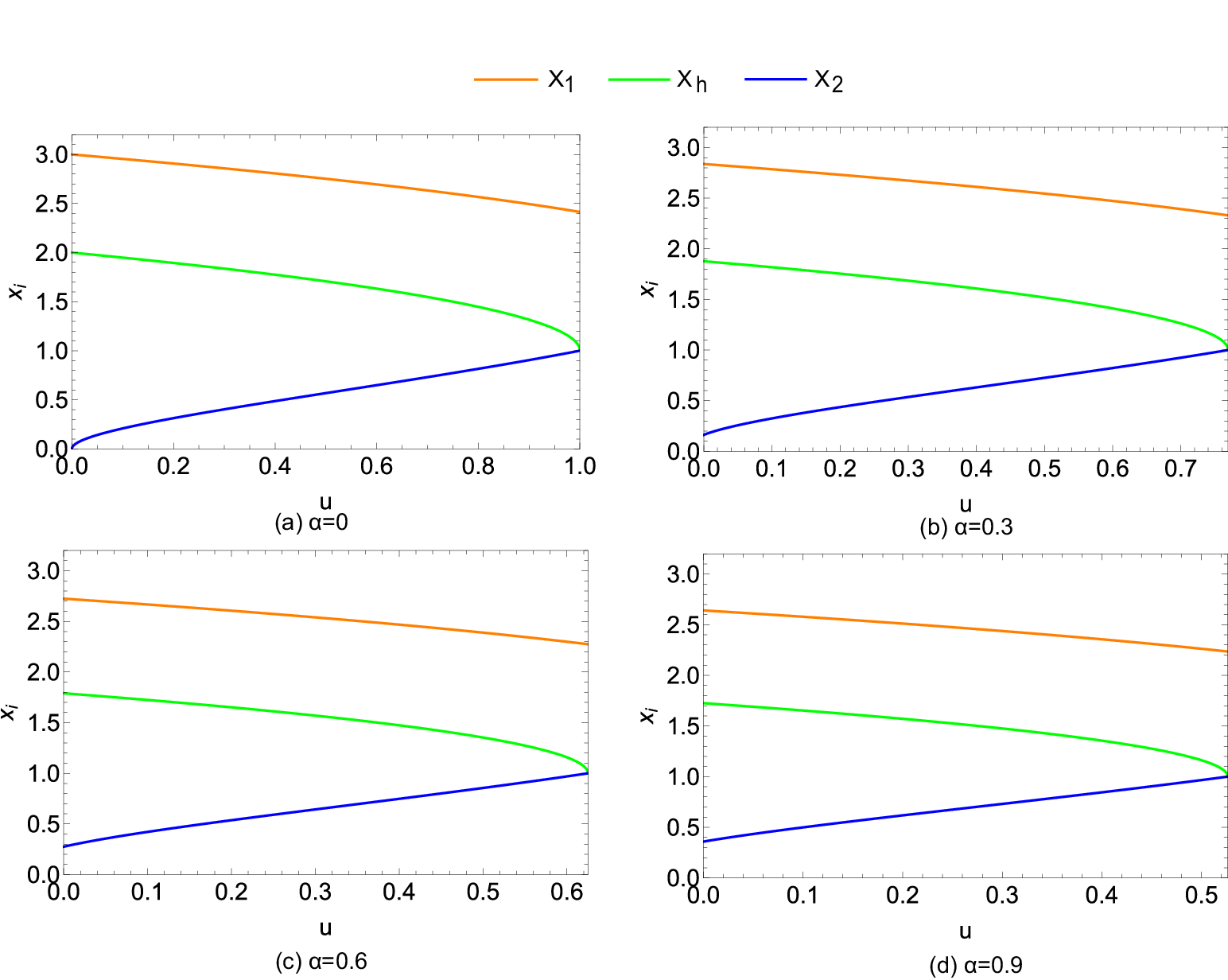}
\caption{The behavior of positive polar photon orbits ($x_1$ and $x_2$) and event horizon $x_h$ versus the rotation parameter $u$ for different deviation parameter $\alpha$.}
\label{fig2}
\end{figure}

According to Eq.~(\ref{eq29}) and Eq.~(\ref{eq30}), one can plot all the polar photon orbits and the event horizon of Kerr-MOG black holes as the functions of the rotation parameter $u$ and deviation  parameter $\alpha$ in Fig.~\ref{fig1}.  It is observed that one photon orbit, $x_1$, exists outside the event horizon $x_h$, while two photon orbits, $x_2$ and $x_3$, lie inside $x_h$. Note that $x_3$ is negative, making it unphysical. Thus, the unphysical roots will be excluded in subsequent analysis. Studies on black hole shadows and gravitational lensing \cite{Cunha2017, Sheoran2018, Liu2024kerr, Kuang2022constraining} primarily focus on $x_1$, as only photon orbits outside the event horizon are observable. To further explore the effects of various parameters on photon orbits around black holes, we provide Fig.~\ref{fig2}.

\vspace{-1.2\baselineskip}
\vspace{-0.8\baselineskip}
Figure~\ref{fig2} illustrates the relationship between the positive photon orbits and the rotation parameter $u$ for various values of the deviation parameter $\alpha$. As depicted in Fig.~\ref{fig2}(a), when $\alpha=0$, the results correspond to the classical Kerr black hole case.	However, as $\alpha$ increases, the photon orbits and event horizon progressively deviate from the classical case.	It is observed that $x_1$ and $x_h$ decrease with increasing $\alpha$, while $x_2$ increases as $\alpha$ grows.	 Furthermore, as $\alpha$ increases, the range of permissible values for $u$ shrinks significantly.	For the extreme case (i.e. $u = 1/\left( 1+ \alpha \right)$), the $x_2$ and $x_h$ overlap together. Therefore, the extreme Kerr-MOG black hole only has one photon orbit, which located out the evert horizon.

\begin{figure}[htbp]
\centering
\includegraphics[width=0.7\linewidth]{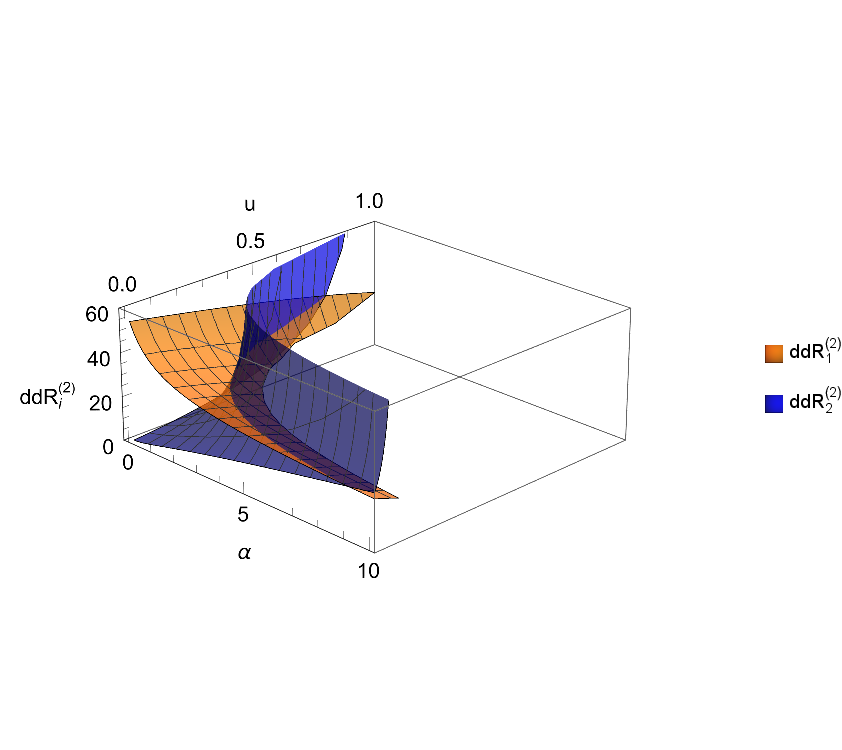}
\vspace{-2.5\baselineskip}
\caption{The ${\text{ddR}}_i^{\left( {2} \right)}$ as functions of $\alpha$ and $u$ for pole solutions $x_i$.}
\label{fig3}
\end{figure}

Next, we analyze the stability of photon orbits, which determines whether photons are repelled from the black hole or fall into it under radial perturbations. According to Ref.~\cite{Cunha2017}, the stability of photon orbits is determined by
\begin{equation}
{\text{ddR}}_i^{\left( {2} \right)} = {\left. {\frac{{{\text{d}^2}R\left( x \right)}}{{\text{d}{x^2}}}} \right|_{x = {x_i}}},
\label{eq30+}
\end{equation}
where $R\left( x \right) = {{R\left( r \right)} \mathord{\left/ {\vphantom {{R\left( r \right)} {{M^4}{E^2}}}} \right. \kern-\nulldelimiterspace} {{M^4}{E^2}}}$. It is well known that ${\text{ddR}}_i^{\left( {2} \right)} \leq 0$ indicates a stable photon orbit, whereas ${\text{ddR}}_i^{\left( {2} \right)} > 0$ represents an unstable photon orbit. By substituting the roots of Eq.~(\ref{eq29}) into Eq.~(\ref{eq30+}), we depict the behavior of ${\text{ddR}}_i^{\left( {2} \right)}$ as functions of $u$ and $\alpha$ in Fig.~\ref{fig3}. The orange and blue surfaces illustrate the stability of photon orbits $x_1$ and $x_2$, respectively. It is observed that ${\text{ddR}}_1^{\left( {2} \right)}$ and ${\text{ddR}}_2^{\left( {2} \right)}$ remain positive, indicating that the polar photon orbits $x_1$ and $x_2$ are unstable under radial perturbations. Since photon orbit $x_2$ resides inside the event horizon, it cannot be observed by detectors at infinity. In contrast, the photon orbit $x_1$ lies outside the event horizon and is observable. Therefore, it is essential to analyze the critical impact parameter of $x_1$.

\vspace{-0.83\baselineskip}
\vspace{-0.7\baselineskip}
The impact parameter is a key physical quantity that describes the motion of particles or light within the gravitational field of black holes.	For rotating black holes, the inclusion of spin and modified gravity parameters introduces asymmetry and complexity to photon trajectories.	Consequently, it is crucial to study the behavior of light photons originating from infinity as they traverse the vicinity of a Kerr-MOG black hole.	As described in Refs.~\cite{Carter1968, Moffat2005}, the critical impact parameter $\beta$ for a four-dimensional axisymmetric spacetime is defined as:
\begin{equation}
\beta = \frac{L}{ME} = \sqrt{\xi^2 + \eta},
\label{eqa31+}
\end{equation}
where $L$ is the angular momentum of photons. Substituting Eq.~(\ref{eq23}) into Eq.~(\ref{eqa31+}), the critical impact parameter of the Kerr-MOG black hole is given by
\begin{equation}
\beta  = \sqrt {\frac{{u{{\left( {1 + x} \right)}^2}\left( {1 + \alpha } \right) + 2x\left[ {2\alpha  + x\left( {{x^2} - 3} \right)\left( {1 + \alpha } \right)} \right]}}{{{{\left( {x - 1} \right)}^2}\left( {1 + \alpha } \right)}}}.
\label{eq36}
\end{equation}
\begin{figure}[htbp]
\centering
\vspace{-1.\baselineskip}
\includegraphics[width=0.6\linewidth]{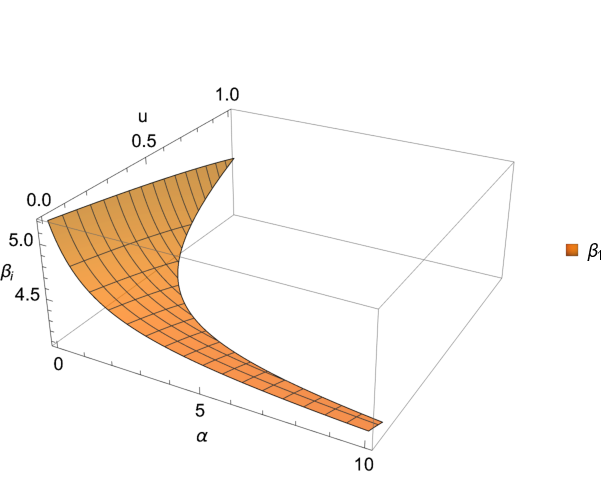}
\caption{The critical impact parameter $\beta_i$ for polar orbits  as functions of $\alpha$ and $u$.}
\label{fig4}
\end{figure}

Figure~\ref{fig4} illustrates the critical impact parameter for photons on the polar orbit $x_1$ of a Kerr-MOG black hole, as derived from Eq.~(\ref{eq36}). The critical impact parameter delineates three possible outcomes for photons near the black hole:	(i) Photons with impact parameters smaller than the critical value $\beta_1$ are absorbed by the black hole;	(ii) Photons with impact parameters equal to $\beta_1$ enter the unstable orbit $x_1$, rotating several times before eventually escaping; (iii) Photons with impact parameters greater than $\beta_1$ are scattered to infinity and can be detected. The surface $\beta_1$ decreases with increasing $\alpha$, indicating that the MOG effect significantly diminishes the black hole's capacity to capture photons. In other words, more photons are scattered to infinity, increasing the likelihood of detection. Consequently, from an observational standpoint, the polar photon ring of a Kerr-MOG black hole should appear brighter compared to a Kerr black hole.

\subsection{The photon orbits of equatorial plane}
\label{sec4-2}
\begin{figure}[htbp]
\centering
\includegraphics[width=0.7\linewidth]{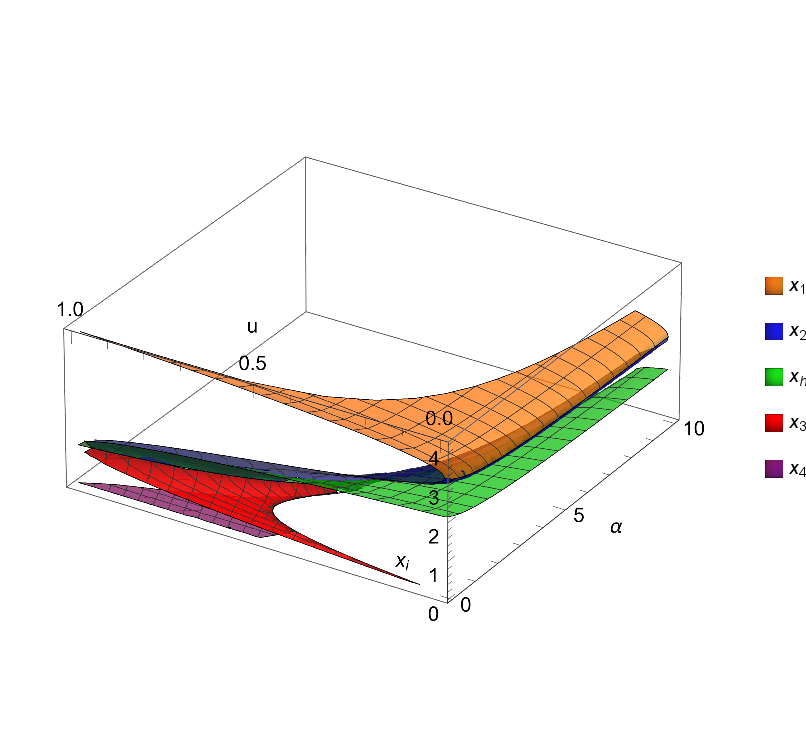}
\vspace{-1.5\baselineskip}
\caption{The equatorial photon orbits $x_i$ and event horizon $x_h$ of Kerr-MOG black hole as the functions of the rotation parameter $u$ and deviation parameter $\alpha$.}
\label{fig5}
\end{figure}

For photon orbits of the equatorial plane, one has $ i = 0 $ or $\pi $, which leads to $ v = 1 $ and $ \mathcal{K} = 0 $. Therefore, Eq.~(\ref{eq28}) is simplified as follows
\begin{equation}
f\left(x\right) =
\left[ 2\alpha - 3x(1 + \alpha) + x^2(1 + \alpha) \right]^2 - 4u \left[ x + 3x\alpha - \alpha(1 + \alpha) \right].
\label{eq31}
\end{equation}
By solving Eq.~(\ref{eq31}), one can obtain four different real roots in the nonextremal case, denoted in descending order as $x_1$, $x_2$, $x_3$ and $x_4$, respectively
Based on these roots, Fig.~\ref{fig5} illustrates the behavior of equatorial photon orbits for various values of the parameters $\alpha$ and $u$.
\vspace{-0.3\baselineskip}

\begin{figure}[htbp]
\centering
\includegraphics[width=0.7 \linewidth]{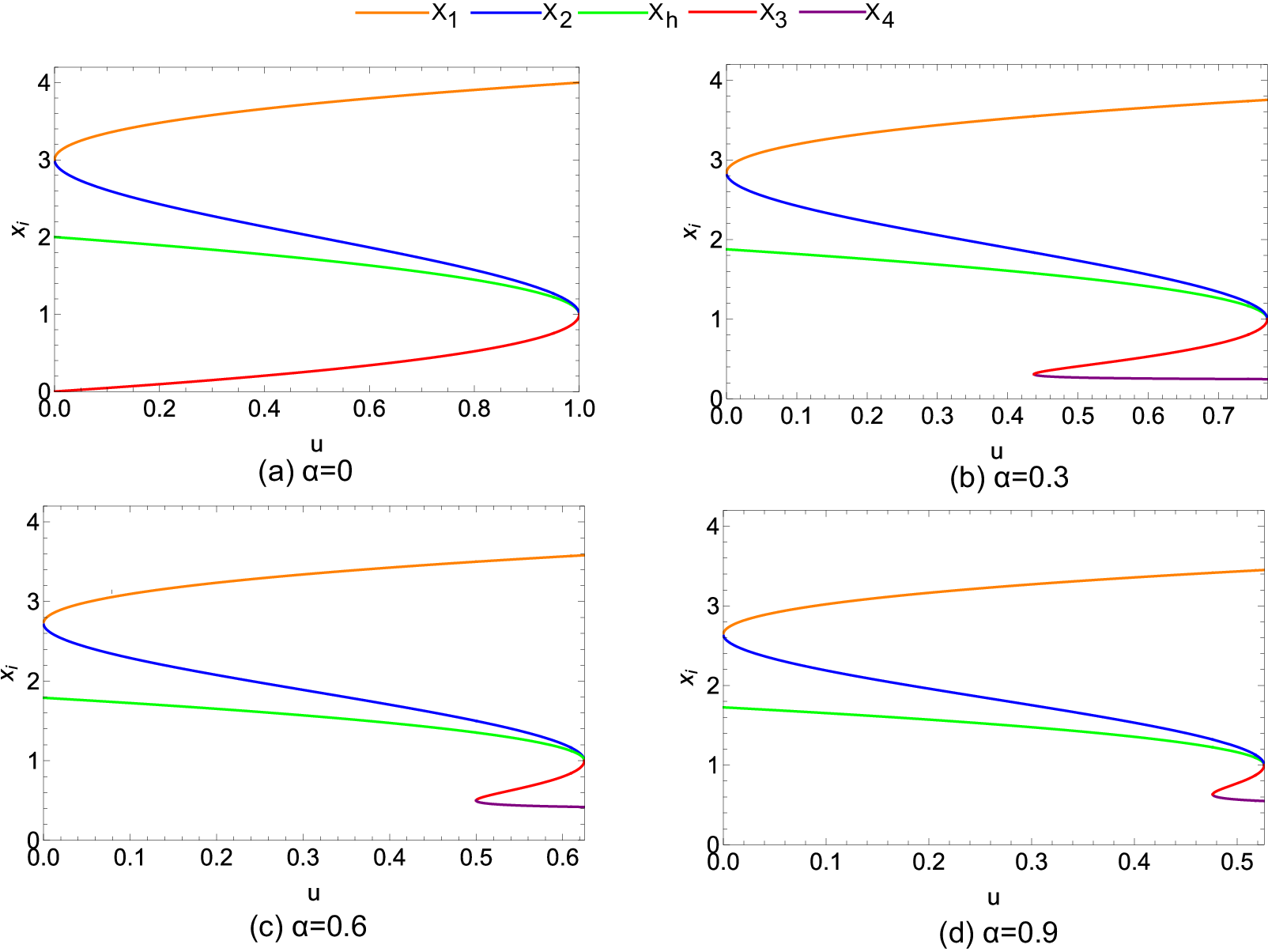}
\caption{The behavior of equatorial photon orbits $x_i$ and event horizon $x_h$ versus the rotation parameter $u$ for different deviation parameter $\alpha$.}
\label{fig6}
\end{figure}

As can be seen in Fig.~\ref{fig5}, one obtains the classical Kerr case for $\alpha = 0$, where there are only three equatorial photon orbits in spacetime, i.e., the retrograde orbits $x_1$, the sequential orbits $x_2$, and the internal orbits $x_3$ ($x_4$ is unphysical since it is negative). The retrograde orbit increases monotonically with $u$, while the prograde orbit decreases monotonically \cite{Teo2003}. The introduction of the deviation parameter $\alpha$ significantly alters the behavior of all equatorial photon orbits. As $\alpha$ increases, both retrograde and prograde orbits have monotonically reduced radii. This distinction offers a potential observational method to differentiate between Kerr-MOG and Kerr black holes by examining photon rings. This phenomenon arises because $\alpha$ reduces the upper limit of the rotation parameter $u$ (see Fig.~\ref{fig6}), thereby decreasing the angular momentum lost by photons in retrograde orbits and the angular momentum gained by photons in prograde orbits. From Fig.~\ref{fig6}(b)-Fig.~\ref{fig6}(d), it is also found that Kerr-MOG has two photon orbits (red curve for $x_3$ and purple curve for $x_4$) within the event horizon, which is never present in the Kerr black hole. Additionally, for the extremal case (i.e. $u = 1/\left( 1+ \alpha \right)$), the $x_2$, $x_3$ and $x_h$ overlap together. Therefore, the extremal Kerr-MOG black hole only has two equatorial photon orbits, one ($x_1$) outside the event horizon and the other ($x_1$) inside the event horizon.

\begin{figure}[htbp]
\centering
\includegraphics[width=0.75\linewidth]{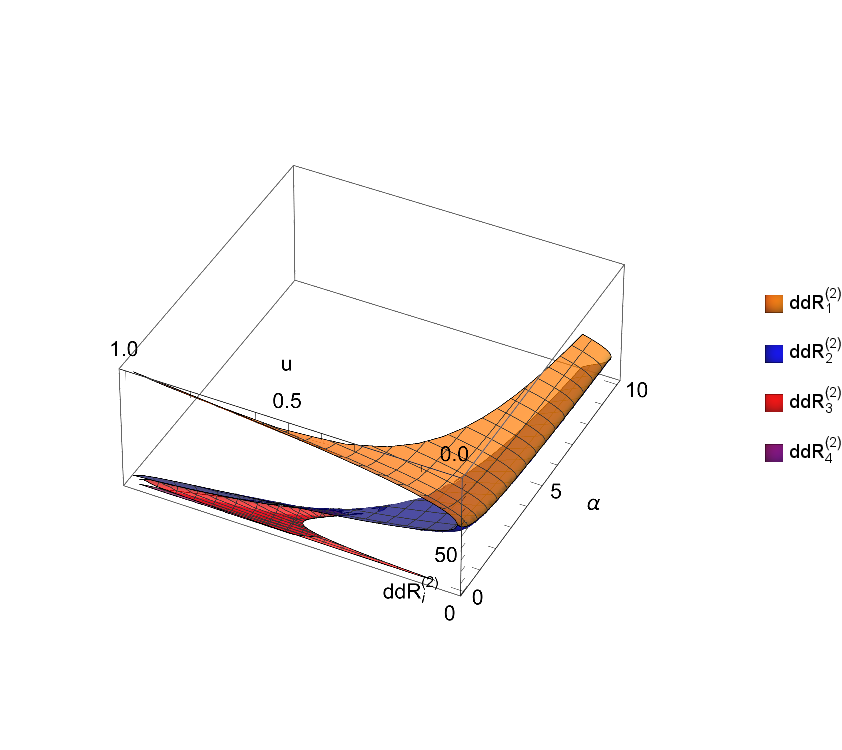}
\vspace{-1.5\baselineskip}
\caption{The ${\text{ddR}}_i^{\left( {2} \right)}$ as functions of $\alpha$ and $u$ for equatorial photon orbits $x_i$.}
\label{fig7}
\end{figure}

Next, we examine the radial stability of equatorial photon orbits using Eq.~(\ref{eq30+}).	As shown in Fig.~\ref{fig7}, all functions ${\text{ddR}}_i^{\left( {2} \right)}>0$, indicating that all equatorial photon orbits are radially unstable.

\begin{figure}[htbp]
\centering
\includegraphics[width=0.65\linewidth]{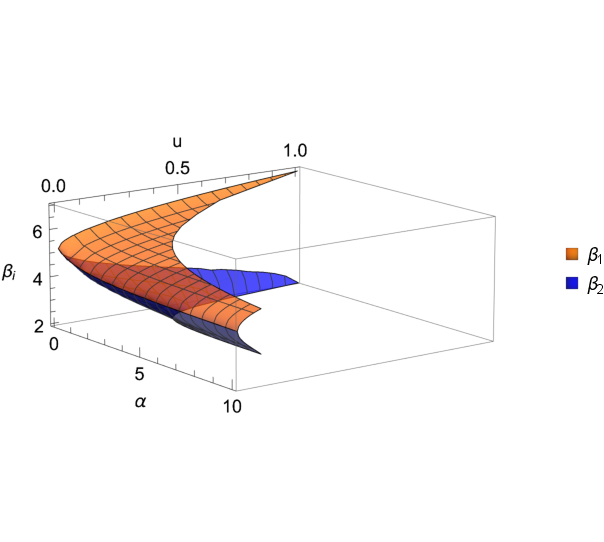}
\vspace{-1.8\baselineskip}
\caption{The critical impact parameters $\beta_i$  for photons in the retrograde and prograde orbits on the equatorial plane of the Kerr-MOG black hole.}
\label{fig8}
\end{figure}

To analyze the behavior of photons around the retrograde and prograde orbits, we calculate the critical impact parameters for photon orbits using Eq.(\ref{eq36}) and illustrate the results in Fig.\ref{fig8}. The orange and blue surfaces correspond to the critical impact parameters $\beta_i$ for photons in retrograde and prograde orbits on the equatorial plane of the Kerr-MOG black hole, respectively.	It is observed that $\beta_1$ for the retrograde orbit increases with the rotation parameter $u$, while $\beta_2$ for the prograde orbit decreases as $u$ increases. This suggests that for rotating black holes, photons in prograde orbits are more likely to scatter to detectors at infinity than those in retrograde orbits, making the photon ring of the prograde orbit appear brighter. Moreover, the deviation parameter $\alpha$ reduces the range of $u$, decreasing the critical impact parameters for both prograde and retrograde orbits. This implies that photon rings in both prograde and retrograde orbits on the equatorial plane are dimmer compared to those of the Kerr black hole.
\vspace{-1.2\baselineskip}

\section{General photon orbits of Kerr-MOG black hole}
\label{sec5}
\begin{figure}[htbp]
\centering
\includegraphics[width=0.65\linewidth]{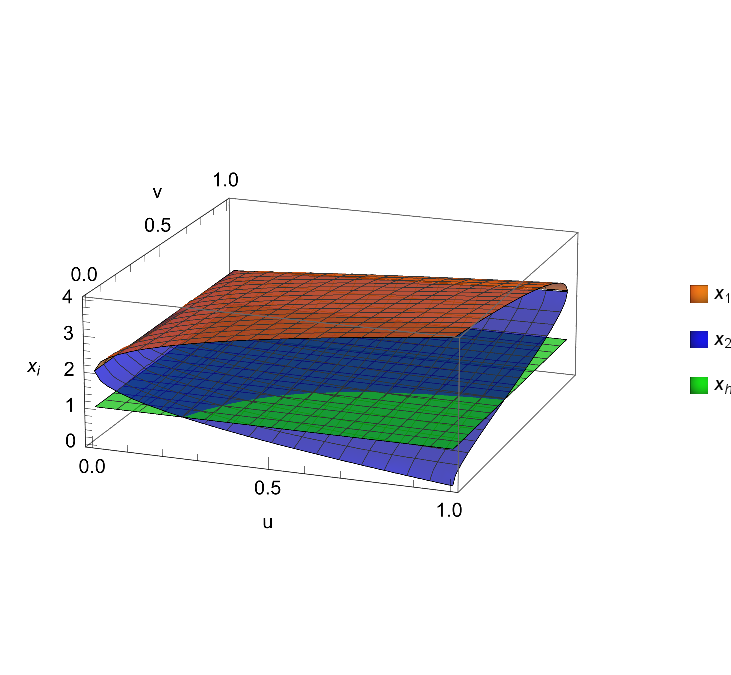}
\vspace{-1.8\baselineskip}
\caption{The general photon general orbits and event horizon of extreme Kerr-MOG black hole as functions of the rotation parameter $u$ and the effective inclination angle $v$.}
\label{fig13}
\end{figure}

In this section, we analyze the most general scenario of spherical photon orbits ($0<v<1$) around a Kerr-MOG black hole. First, we examine the extreme case of a Kerr-MOG black hole, where $u = {1 \mathord{\left/ {\vphantom {1 {\left( {1 + \alpha } \right)}}} \right. \kern-\nulldelimiterspace} {\left( {1 + \alpha } \right)}}$ and $0 < u < 1$. Substituting $\alpha$ in Eq.~(\ref{eq28}) with $u$, one obtains
\begin{align}
{\left( {x - 1} \right)^2}{\mathcal{P}_4}(x) = 0,
\label{eq35}
\end{align}
where
\begin{align}
{\mathcal{P}_4}\left( x \right) = {u^2}v + {\left( {x - 2} \right)^2}{x^2} + 2ux\left[ {\left( {2 + x} \right)v - 2x} \right].
\label{eq36}
\end{align}
It is easily found that the double root in $\left(x - 1 \right)^2$ of Eq.~(\ref{eq25}) corresponds to the event horizon radius, given by $x = 1$. According to the Descartes' rule of signs, the quartic ${\mathcal{P}_4}\left( x \right) = 0$ can have at most two positive real roots, denoted as $x_1$ and $x_2$. The analytical expressions for these roots are complex, so their relationships with the parameters $v$ and rotation parameter $u$ are illustrated in Fig.~\ref{fig13}. For a slowly rotating extremal Kerr-MOG black hole ($0 < u < 0.25$), two photon orbits always exist outside the event horizon ($x_h$), independent of the effective inclination angle.	However, for a rapidly rotating extremal Kerr-MOG black hole ($0.25 < u < 1$), there exists a critical inclination angle $v_{cr} = {{\left( {4u - 1} \right)} \mathord{\left/ {\vphantom {{\left( {4u - 1} \right)} {u\left( {6 + u} \right)}}} \right. \kern-\nulldelimiterspace} {u\left( {6 + u} \right)}}$. When $v < v_{cr}$, only a single retrograde photon orbit ($x_2$) exists outside the event horizon.	Conversely, when $v>v_{cr}$, there are two photon orbits (prograde photon orbit $x_1$ and retrograde photon orbit $x_2$) are present outside the event horizon.		

Next, we turn our attention to the non-extremal case.  It is well-established that general algebraic equations of degree higher than quartic do not have analytic solutions. Consequently, the sextic polynomial~(\ref{eq28}) cannot be solved directly.	 Numerical methods could be employed to determine the radius of the orbit, but they yield $x=\left(u, \alpha, v \right)$ rather than the desired $x=\left(u, \alpha \right)$.  According to  Ref.~\cite{Hod2013}, it is found a critical inclination angle  ${v_{cr}} = {v_{cr}}\left( {u,\alpha } \right)$ in the $\left(u, \alpha, v \right)$ parameter space, below this critical inclination angle there are four real roots and above it there are two real roots. To determine this critical inclination angle, the parameter $v$ must be expressed in terms of $u$ and $\alpha$ as	
\begin{align}
v = \frac{\Xi }{u},
\label{eq33}
\end{align}
where
\begin{align}
\Xi & =  {\left( {w - 1} \right)^2}\left[ {3{w^2}\left( {5 + 6\alpha  + {\alpha ^2}} \right) - w\left( {3 + \alpha  - 2{\alpha ^2}} \right) - 4\left( {3 + 2\alpha } \right)} \right]
\nonumber \\
&- u\left( {1 + \alpha } \right)\left[ {3w\left( {1 + \alpha } \right) - 4\left( {3 + 2\alpha } \right)} \right]/w\left[ {u + 6{{\left( {w - 1} \right)}^2}} \right]\left( {w - 2} \right){\left( {1 + \alpha } \right)^2},
\label{eq33-1}
\end{align}
\begin{align}
w = \frac{{{{\left[ {\left( {1 - u\alpha  - u} \right){{\left( {1 + \alpha } \right)}^2}} \right]}^{\frac{1}{3}}}}}{{1 + \alpha }}.
\label{eq33-3}
\end{align}
At the critical point, one has
\begin{align}
\label{eq33-2}
{\Xi _{cr}} = \frac{{{{\left( {1 - w} \right)}^3}\left[ {3w\left( {1 + \alpha } \right) + \alpha  - 3} \right]}}{{w\left[ {12 + \left( {w - 6} \right)w} \right]\left( {1 + \alpha } \right) - 7 - 6\alpha }}.
\end{align}

\begin{figure}[htbp]
\centering
\includegraphics[width=0.55\linewidth]{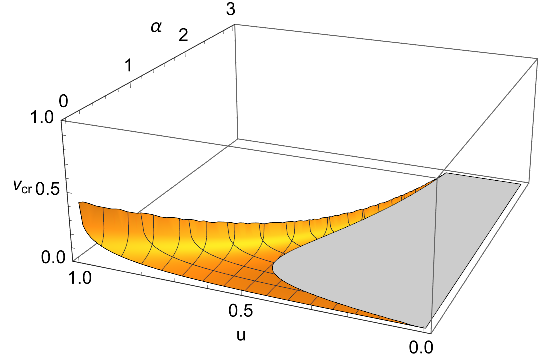}
\caption{The critical inclination angle as the functions of the rotation parameter $u$ and $\alpha$.}
\label{fig8+}
\end{figure}

Using Eq.(\ref{eq33-1}) and Eq.(\ref{eq33-2}), the variation of the critical inclination angle $v_{cr}$ with $\alpha$ and $u$ is illustrated in Fig.~\ref{fig8+}. It is evident that for $\alpha \ne 0$, part of $v$ (indicated by the gray area) becomes less than $0$, violating the condition $0 < v < 1$. Therefore, to analyze general photon orbits of the Kerr-MOG black hole in the parameter space $\left(\alpha, u \right)$, it is essential to impose constraints on $\alpha$ and $u$. By solving the inequality $v > 0$, a new lower bound of $u$ can be obtained as ${{u > \left( {81 + 18\alpha  + {\alpha ^2}} \right)\alpha } \mathord{\left/
 {\vphantom {{u \geqslant \left( {81 + 18\alpha  + {\alpha ^2}} \right)\alpha } {27\left[ {1 + \alpha \left( {3 + 3\alpha  + {\alpha ^2}} \right)} \right]}}} \right.
 \kern-\nulldelimiterspace} {27\left[ {1 + \alpha \left( {3 + 3\alpha + {\alpha ^2}} \right)} \right]}}$. Moreover, to prevent a naked singularity in spacetime, the rotation parameter $u$ must satisfy the constraint $u < {1 \mathord{\left/ {\vphantom {1 {1 + \alpha }}} \right. \kern-\nulldelimiterspace} ({1 + \alpha})}$, as dictated by Eq.~(\ref{eq30}). Consequently, the ranges of $u$ and $\alpha$ are defined as ${{\alpha \left( {81 + 18\alpha + {\alpha ^2}} \right)} \mathord{\left/ {\vphantom {{\alpha \left( {81 + 18\alpha + {\alpha ^2}} \right)} {27\left[ {1 + \alpha \left( {3 + 3\alpha + {\alpha ^2}} \right)} \right]}}} \right. \kern-\nulldelimiterspace} {27\left[ {1 + \alpha \left( {3 + 3\alpha + {\alpha ^2}} \right)} \right]}} < u < {1 \mathord{\left/ {\vphantom {1 {\left( {1 + \alpha } \right)}}} \right. \kern-\nulldelimiterspace} {\left( {1 + \alpha } \right)}}$ and $0 < \alpha < 3$. Obviously, when $\alpha=0$, the result reduces to the Kerr case. Within the region $0 < v < 1$, there are four roots below the critical surface $v_{cr}$, consistent with the results on the equatorial plane, whereas above this surface, only two roots exist, aligning with the results on the polar plane.	

\begin{figure}[htbp]
\centering
\includegraphics[width=0.7\linewidth]{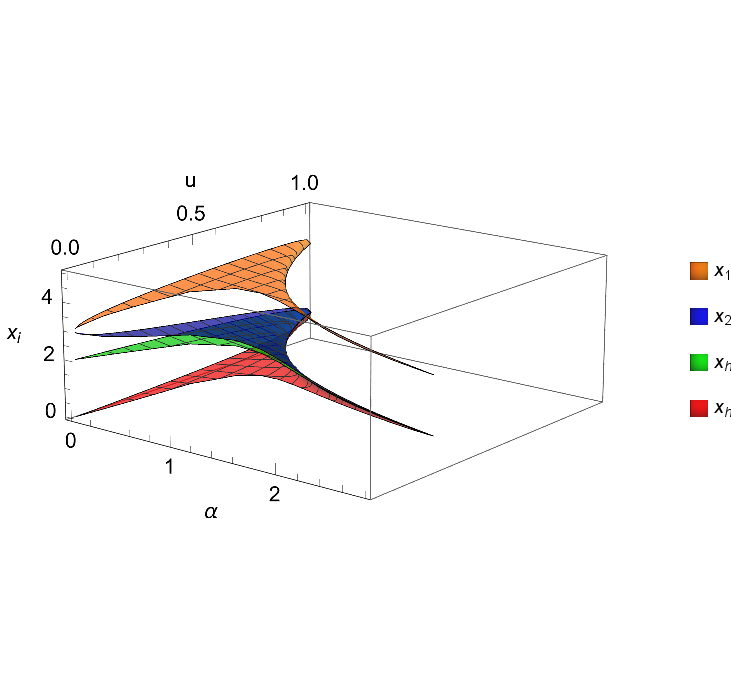}
\vspace{-1.5\baselineskip}
\caption{The general photon orbits $x_i$ and event horizon $x_h$ of Kerr-MOG black hole as functions of the rotation parameter $u$ and deviation parameter $\alpha$.}
\label{fig9}
\end{figure}

Substituting Eq.(\ref{eq33-2}) into Eq.(\ref{eq33}) yields the critical inclination angle. At this angle, the sextic~(\ref{eq28}) can be expressed as the product of a quartic $\mathcal{P}_4\left(x\right)$ and a quadratic $\mathcal{P}_2\left(x\right)$ as follows
\begin{align}
f\left(x\right) =
\mathcal{P}_4\left(x\right) \mathcal{P}_2\left(x\right) = \left(x^4 + A_1 x^3 + A_2 x^2 + A_3 x + A_4 \right)\left(x + A_5\right)^2,
\label{eq32}
\end{align}
where $A_1$, $A_2$, $A_3$, $A_4$, and $A_5$ denote the coefficients  consisting only of the rotation parameter $u$ and the deviation parameter $\alpha$.  By comparing with Eq.~(\ref{eq28}) with Eq.~(\ref{eq32}), the coefficients are given by
\begin{subequations}
\label{eq35}
\begin{align}
A_1 &= -4 - 2w,
\\
A_2 &=(-6+3w^{5}a^{2}+3w^{5}+6w^{5}a-6wa+44w^{3}a+20w^{3}a^{2}+24w^{3}
\nonumber \\
&-18w-18a^2w^4-18w^4-36w^4a+15w^2+33w^2a+12a^2w+18a^2w^2
\nonumber \\
&-22a^2-32a)/(-6+3w^5a^2+3w^5+6w^5a-6wa+44w^3a+20w^3a^2
\nonumber \\
&+24w^3-18w-18a^2w^4-18w^4-36w^4a+15w^2+33w^2a+12a^2w
\nonumber \\
&+18a^{2}w^{2}-22a^{2}-32a),
\\
A_3&=2(-6+3w^{5}\alpha^{2}+3w^{5}+6w^{5}\alpha+17w\alpha+4w^{3}\alpha-2w^{3}\alpha^{2}+6w^{3}+9w
\nonumber \\
&-5a^2w^4-9w^4-14w^4a-3w^2-9w^2a+4a^2w-6a^2w^2+2a^2-4d)/(-7
\nonumber \\
&+2w^{3}a-6w^{2}+24wa+12a^{2}w-6a^{2}w^{2}+w^{3}a^{2}+12w+w^{3}-6a^{2}-13a
\nonumber \\
&-12w^{2}a),
\\
A_4 &=(-3+3w^5a^2+3w^5+6w^5a+2wa+2w^3a-w^3a^2+3w^3+6w
\nonumber \\
&-2a^2w^4-6w^4-8w^4a-3w^2-3w^2a+a)/(-7+2w^3a-6w^2
\nonumber \\
&+24wa+12a^2w-6a^2w^2+w^3a^2+12w+w^3-6a^2-13a-12w^2a),
\\
A_5 &= w-1.
\end{align}
\end{subequations}
According to Eq.~(\ref{eq35}), the roots of Eq.~(\ref{eq28}) can be obtained. For the quartic part $\mathcal{P}_4\left(x\right)$, there are two real roots and two complex roots. For the quadratic part $\mathcal{P}_2\left(x\right)$, there exists a single double root expressed as $x = 1 - w$. Due to the length and complexity of these expressions, they are not explicitly presented here. Instead, numerical methods are employed to depict the behavior of these roots as functions of $u$ and $\alpha$, shown in Fig.~\ref{fig9}. It is seen that two real roots (i.e., $x_1$ and $x_2$) from  $\mathcal{P}_4\left(x\right)$ part are located outside the event horizon (green surface). The orange surface for $x_1$ is defined as the retrograde orbit since the photons in it move in the opposite direction to the black hole rotation requiring angular momentum in a larger orbit, while blue surface for $x_2$ is represented the prograde orbit based on the fact that the photons in it move in the same direction as the black hole's rotation. The multiple root from $\mathcal{P}_2\left(x\right)$ exists inside the event horizon.

\begin{figure}[htbp]
\centering
\includegraphics[width=0.7\linewidth]{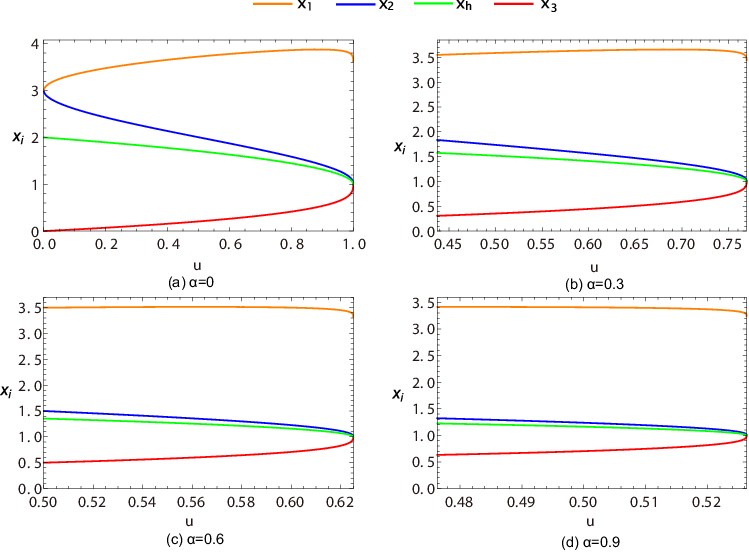}
\caption{The behavior of general photon orbits $x_i$ and event horizon $x_h$ versus the rotation parameter $u$ for different deviation parameter $\alpha$.}
\label{fig10}
\end{figure}

To facilitate the discussion of how the deviation parameter affects general photon orbits, we plot the behavior of $x_i$ as a function of $u$ for different $\alpha$, derived from the roots of $\mathcal{P}_4\left(x\right)$ and $\mathcal{P}_2\left(x\right)$. Figure~\ref{fig10}(a) shows the general photon orbits of a classical Kerr black hole when $\alpha=0$. As $u$ increases, $x_2$ and $x_h$ decrease monotonically, whereas $x_3$ increases monotonically. Interestingly, $x_1$ initially increases but then decreases as it approaches $1$. However, as $\alpha$ increases, the range of $u$ gradually narrows, and the variations in all photon orbits and the event horizon become less pronounced. Specifically, for large $\alpha$, the prograde photon orbits remain almost unaffected by $u$, which significantly differs from the behavior in Kerr black holes. Additionally, the prograde photon orbits are located very close to the event horizon, enabling the analysis of black hole event horizon properties through observations of the prograde photon orbit.	

\begin{figure}[htbp]
\centering
\includegraphics[width=0.65\linewidth]{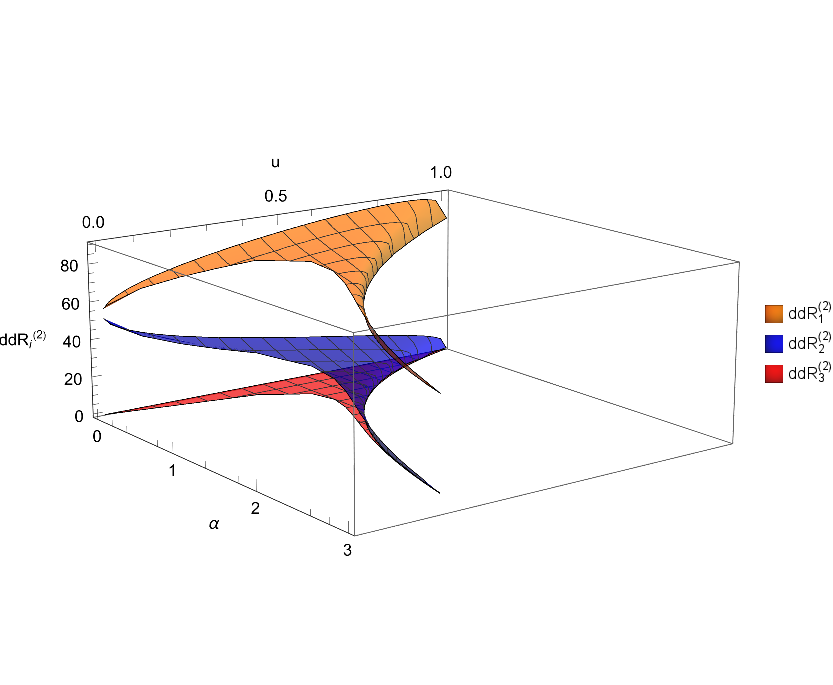}
\vspace{-1.5\baselineskip}
\caption{The ${\text{ddR}}_i^{\left( {2} \right)}$ as functions of $\alpha$ and $u$ for general photon orbits $x_i$.}
\label{fig11}
\end{figure}

\begin{figure}[htbp]
\centering
\includegraphics[width=0.7\linewidth]{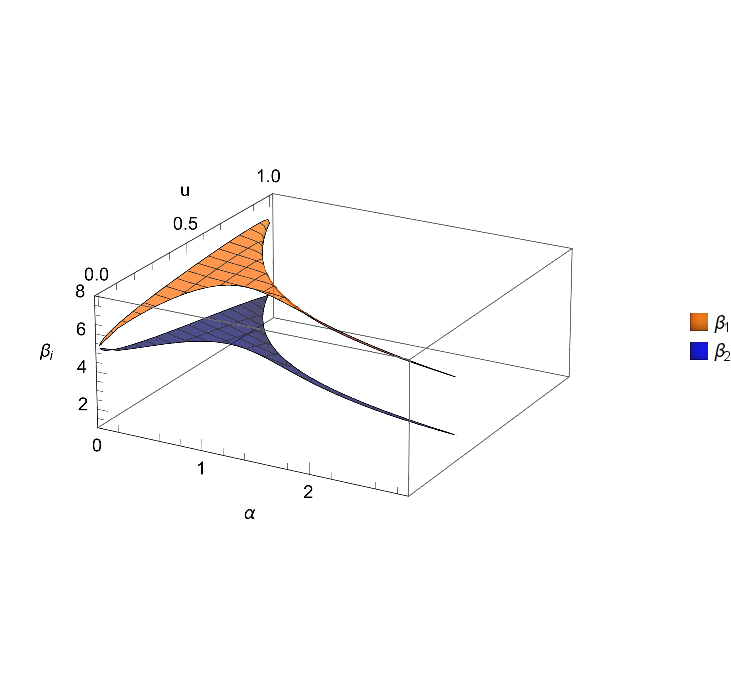}
\vspace{-2.5\baselineskip}
\caption{The critical impact parameters $\beta_i$  of photons on the general plane of the retrograde and prograde orbits as the functions of $\alpha$ and $u$.}
\label{fig12}
\end{figure}

Finally, we analyze the radial stability $\text{ddR}^{(2)}_i$ and the critical impact parameters $\beta_i$ for all general photon orbits using Eq.~(\ref{eq30}) and Eq.~(\ref{eq36}). Fig.~\ref{fig11} shows that $\text{ddR}^{(2)}_i > 0$, indicating that all general orbits are radially unstable.	Moreover, Fig.~\ref{fig12} illustrates that the impact parameter of general photon orbits are very similar to those on the equatorial plane.	Thus, it can be concluded that the rotation parameter increases $\beta_1$ for retrograde orbits while decreasing $\beta_2$ for prograde orbits, resulting in a significant difference in photon scattering probabilities.	Additionally, $\alpha$ significantly restricts the range of rotation parameters and reduces the critical impact parameters for both retrograde and prograde orbits.	Therefore, the brightness of general photon orbits in Kerr-MOG black holes is lower than that in Kerr black holes.	


\section{Conclusion and discussion}
\label{sec6}
Studying the photon orbits of black holes is of great significance for understanding the structure and properties of the gravitational field of black holes. In this paper, we investigated the photon orbits around Kerr-MOG black holes, focusing on the effects of the rotation parameter $u$, deviation parameter $\alpha$, and effective inclination angle $v$. By solving the Hamilton-Jacobi equation, we derived a sixth-order polynomial~(\ref{eq28}). Then, we examined the polar, equatorial, and general photon orbits and analyzed the effects of key parameters using numerical methods.

In the polar case ($v=1$), two valid photon orbits were identified: $x_1$ lies outside the event horizon $x_h$, while $x_2$ resides inside. Both the radii of  $x_1$ and $x_h$ decrease with increasing deviation parameter $\alpha$ and rotation parameter $u$, whereas the radius of $x_2$ increases with these parameters.	These results indicate that $\alpha$ modifies gravitational effects around the black hole, distinguishing Kerr-MOG black holes from classical Kerr black holes. The stability analysis reveals that polar orbits are radially unstable. Furthermore, the critical impact parameter $\beta_i$ decreases with increasing $\alpha$, suggesting that photons with smaller angular momentum are more likely to be captured by the black hole.	

In the equatorial case ($v=0$), four valid photon orbits exist: $x_1$ (retrograde) and $x_2$ (prograde) are outside the event horizon, while $x_3$ and $x_4$ are inside.	As $\alpha$ increases, radius of the prograde orbit and retrograde orbit  expands. Furthermore, as the rotation parameter $u$ increases, the radii of the event horizon and prograde orbit decrease, while that of the retrograde orbit increases.	Like polar orbits, all the equatorial orbits are radially unstable. Additionally, the critical impact parameter for prograde orbits decreases with increasing $u$, indicating that faster black hole rotation enhances photon capture on these trajectories. Conversely, the retrograde critical impact parameter increases with $u$, requiring photons with higher angular momentum to sustain their orbits. These behaviors emphasize the sensitivity of equatorial orbits to $\alpha$, providing a clear distinction between Kerr-MOG and Kerr black holes.

In the general case ($0 < v < 1$), it is found that a slowly rotating extremal Kerr-MOG black hole has two general photon orbits outside the event horizon, whereas a rapidly rotating extremal Kerr-MOG black hole has only one such orbit. For non-extremal Kerr-MOG black holes, a critical inclination angle $v_{cr}$ exists in the parameter space $\left(v, u, \alpha \right)$. Below $v_{cr}$, four photon orbits exist, whereas above $v_{cr}$, two are present.	At $v_{cr}$, three solutions are identified: two photon orbits outside and one inside the event horizon. Meanwhile, we find that all general orbits of the Kerr-MOG black hole are radially unstable. The impact parameter decreases with increasing $\alpha$. Thus, the brightness of the Kerr-MOG black hole is suggested to be higher than that of the Kerr black hole.	

\bibliography{references}    

\end{document}